\documentclass{appolb}
\usepackage{graphicx}
\usepackage{cite}
\usepackage{bm}

\begin{document}
\title{Bound-Free Pair Production Mechanism at the NICA Accelerator Complex
}
\author{Melek YILMAZ ~\c{S}ENG\"{u}L
\address{Atakent Mahallesi, 3.Etap, 34303 Halkal{\i}-K\"{u}\c{c}\"{u}k\c{c}ekmece \.{I}stanbul,TURKEY}
\address{melekaurora@yahoo.com}
}
\maketitle
\begin{abstract}

In this work, bound-free pair production (BFPP) cross section calculations are done for NICA collider. In the BFPP mechanisms, after free pair production process, free electron is captured by one of the ions. This causes to the decrement of the ion from the beam, also diminish the intensity of the beam and changes the beam lifetime. BFPP cross section calculations are done for $p-Au$ and $p-Bi$ collisions for NICA collider and it is seen that the obtained results are in the range of a few $mbarns$. In this work, we used the lowest-order perturbation theory and our calculations are done in QED framework.

\end{abstract}
  
\section{Introduction}

In this work, we have studied on bound-free electron-positron pair production mechanism in which electron is captured by the colliding ions

\begin{eqnarray}
\label{2}
   Z_x+Z_y &\rightarrow& (Z_x+e^-)_{1s_{1/2}}\:+\: Z_y \:+\: e^+
\end{eqnarray}

This mechanism is important in the collisions of heavy ions. BFPP cross section at RHIC and LHC energy levels is in the range of hundereds of barns and is smaller than free pair production cross section. However, the calculation of this cross section process is significant for the detection of the reducement of the beam intensity and this process leads to a seperate beam of one-electron ions that impinges the beam pipe about $140 m$ away from the interaction point \cite{1,10}.

The Nuclotron-based Ion Collider fAcility (NICA) Project is in the improvement process at JINR (Dubna). The main aims of the NICA project are to supply colliding beams for experimental studies of both hot and dense strongly interacting baryonic matter and spin physics. The first program will include the running of heavy-ion mode in the energy interval of $\sqrt{s_{NN}}= 4-11 GeV$ for $Au$ nuclei. In this stage part of the project, firstly the fixed target experiments with the heavy ion beam will be deducted from the Nuclotron at kinetic energies up to 4,5 GeV/u. The polarized beam mode is aimed to be studied in the energy interval of $\sqrt{s_{NN}}= 12-27 GeV$ (protons). The detailed information about the NICA can be found in \cite{11}.

The aim of this work is to investigate and to examine the importance of the BFPP mechanism at NICA collider which is under construction. We will consider the proton-nucleus collision with the proton energy up to $12 GeV$. To give reliable contributions to the future experiment at NICA collider, we applied our previous BFPP mechanism calculations to the proton-ion collisions that worked well when we compared our cross section results with \cite{12}.

In the Formalism Part, we gave the detailed information about how we applied our BFPP mechanism calculations for the proton-nucleus collisions of the NICA collider parameters.

In Results and Discussions Part, we gave our proton-nucleus collisions cross section results for the NICA collider and compared our calculations with \cite{12}.

We concluded in Conclusions Part.

\section{Formalism}

In this study, we have calculated the BFPP cross section of the proton-ion collision for NICA collider. We did our calculations by using the semi-classical approximation and we reached the exact results by using the Monte Carlo method. We obtained BFPP cross section in the framework of QED with the help of the lowest-order perturbation theory. Detailed calculations about the BFPP cross section process can be examined in our previous papers \cite{9,13,14}.

In our previous works, we used the symmetric collisons. However, our method can be implemented to asymmetric collisions such as proton-ion collisions. The colliding proton energy is $10GeV$, it is comparable with the $Au$ and $Bi$ and rest masses of the proton and nucleon are not very significiant and this justify to apply the system in the center of momentum frame. The detail formulation can be found in \cite{13}.

We did our calculations by using the semi-classical approximation and we assumed that the positron goes on a straight way. We used Sommerfeld-Maue wave-function for the positron,

\begin{eqnarray}\label{6}
   \Psi^{(+)}_{q} & = & 
   N_{+}\left[e^{i\mathbf{q}\cdot\mathbf{r}}\,
   \textbf{u}^{(+)}_{\sigma_q} \right],
\end{eqnarray}

where $\textbf{u}^{(+)}_{\sigma_q}$ shows the spinor structure.

For the representation of the electron, we used the Darwin wave-function,

\begin{eqnarray}\label{8}
   \Psi^{(-)}(\vec{r}) & = &
   \left( 1-\frac{i}{2m} \, \vec{\alpha}\cdot\vec{\nabla}\right) \:
   \textbf{u} \, \frac{1}{\sqrt{\pi}} \left( \frac{Z}{a_H} \right)^{3/2} \, e^{-Zr/a_H} ,
\end{eqnarray}

Using these wave-functions that are represented above, the direct Feynman diagram can be presented as:
\begin{eqnarray}\label{e21}
   \left\langle\Psi^{(-)}\left|\chi_{pi}\right|\Psi^{(+)}_q\right\rangle \nonumber 
 & = & 
   \frac{iN_{+}}{2\beta} \, \frac{1}{\sqrt{\pi}} \, 
   \left(\frac{Z}{a_H}\right)^{3/2} \,
   \int \frac{d^2p_\bot}{(2\pi)^2} \, 
   e^{i(\mathbf{p}_\bot-\frac{\mathbf{q}_\bot}{2})\cdot\mathbf{b}} \,
   C(\mathbf{p}_\bot,\mathbf{k}_\bot,\beta) \, ,
\end{eqnarray}

where $b$ represents the impact parameter of the proton-ion collision, and $C(\mathbf{p}_\bot,\mathbf{k}_\bot,\beta)$ function expresses the scalar part of the field correlated with the ion $i$ and the proton $p$ in momentum space and transverse momentum components. Detailed information about these functions can be reached in \cite{9,13,14}.

Using the amplitudes for the direct $\chi_{pi}$  and crossed $\chi_{ip}$  diagram, the BFPP cross section in collisions of the proton-ion collision can be expressed as

\begin{eqnarray}\label{9}
   \sigma_{BFPP} & = & 
   \int d^2b\sum_{q<0}\left|\left\langle 
   \Psi^{(-)}\left|\chi_{pi}+\chi_{ip}\right|\Psi^{(+)}_q\right\rangle\right|^2,
\end{eqnarray}

that are the direct and crossed terms summation. These calculations have been accomplished for the BFPP cross sections in relativistic collisions of the proton-ion collision.

For the symmetric heavy ion collisions with charge $Z$, the BFPP cross-section results increase with $\sigma_{BFPP}\approx(Z\alpha)^{8}ln(\gamma)$. The behaviour of the BFPP cross section changes as $\sigma_{BFPP}\approx(Z_{i}\alpha)^{6}(Z_{p}\alpha)^{2}ln(\gamma)$ in an proton-ion collisions, where  $Z_{i}$ expresses the heavy ion and $Z_{p}$ expresses the projectile proton.

\section{Results and Discussions}

\begin{figure}[htb]
\centerline{%
\includegraphics[width=12.5cm]{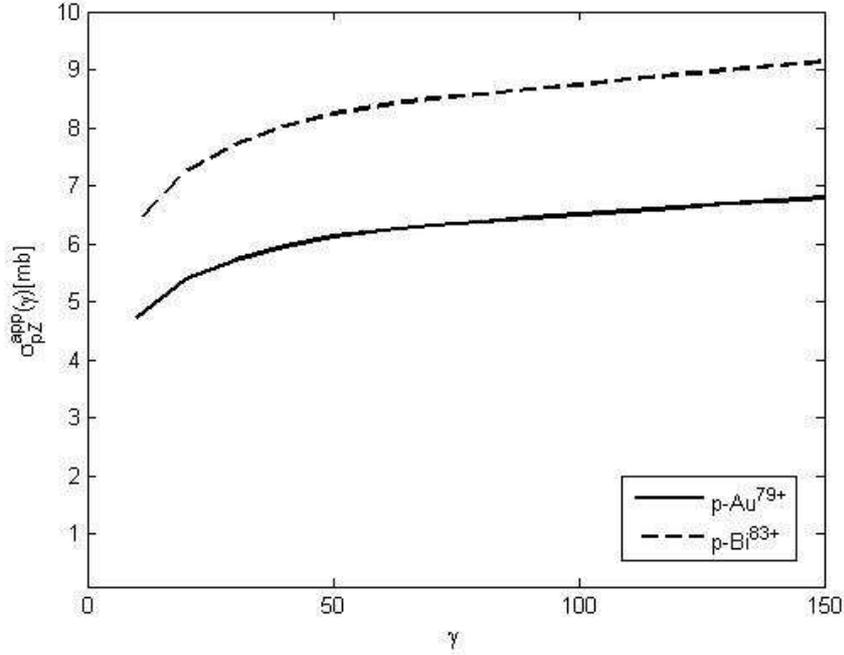}}
\caption{Cross sections ($mbarn$) for $p-Au$ and $p-Bi$ collisions dependent on $\gamma$.}
\label{1}
\end{figure}

$p-Au$ and $p-Bi$ collisions cross section results are seen in Fig.~\ref{1}. BFPP cross section results in $p-Au$ and $p-Bi$ collisions with proton energy 10 GeV and the ion energy 5 $GeV/nucleon$ ($\gamma=100$) at NICA collider are equal to 6,5 $mbarn$ and 8,7 $mbarn$ in our work, respectively. These BFPP cross section results in $p-Au$ and $p-Bi$ collisions at NICA collider obtained in \cite{12} are approximately equal to 4,5 $mbarn$ and 6 $mbarn$, respectively. When we compare our BFPP cross section results for p-Au and p-Bi collisions at the NICA collider with BFPP cross section results obtained in \cite{12}, it is seen that our cross section results are approximately 40 ~\%{} higher than that of in [12]. In \cite{12}, to calculate the $p-Au$ and $p-Bi$ collisions cross section results, equivalent photon approximation is used with the modification of the Sauter approximation. Modifications are done for the NICA energy region. At lower energies and the asymmetric collisions, Eq.~[17] given in \cite{12} does not work. In \cite{12}, new approximation is calculated for NICA collisions and its asymptotic form is identical to Eq.~[17] given in \cite{12}. The advantage of our method is that we can use our approximate BFPP cross section expression for the lower energy of the colliding nuclei and also to the asymmetric collision as proton-nucleus.

In Fig.~\ref{1}, we showed the cross sections for $p-Au$ and $p-Bi$ collisions dependent on $\gamma$. The behaviour of our graph is similar to that of the graph in Fig. 4 given in \cite{12}. However, our cross section results are a few mbarn higher than the cross section results obtained in \cite{12}. The reason for the difference can be explained as follows; we used the more general cross section formula that can be applied in every energy range and also the asymmetric collisions.

We reached BFPP cross section results in the center-of-momentum frame. $\gamma=1/\sqrt{1-v^{2}}=E/m_{0}$  is the relationship between $\gamma$(Lorentz factor) and the collider energy per nucleon $E/A$ which is in $GeV$ unit, in this equation $m_{0}$ represents the mass of the nucleons.

\section{Conclusions}

In the present paper, we did BFPP cross section calculations for NICA collider. The BFPP cross section results for $p-Au$ and $p-Bi$ collisions with the ion energy 5 $GeV/nucleon$ and the proton energy 10 $GeV$ are in the order of magnitude 6- 8 $mbarn$ at NICA collider. We also compared our cross section value with the cross section value obtained in \cite{12}. The results are close to each other. BFPP process leads to a decrease in the intensity of colliding beams. On the other hand, in this work, the calculated cross-section results are very smaller than the cross-section results reached for hadronic interactions, as a result there is no decrease in the intensity of colliding beams. In our calculations, we did not add the effects of the excited atomic states to the capture cross section results that change approximately 10 ~\%{}. Also, with the cross section calculation of $p$-ion collisions, we hope to give contribution to the new building NICA collider.

\end{document}